\documentclass[twocolumn,showpacs,aps,prd,superscriptaddress,floatfix]{revtex4}

\usepackage{graphicx}
\usepackage{dcolumn}
\usepackage{amsmath}
\usepackage{epsfig}
\usepackage{rotating}

\RequirePackage{xspace}

\usepackage{relsize}
\def\babar{\mbox{\slshape B\kern-0.1em{\smaller A}\kern-0.1em
    B\kern-0.1em{\smaller A\kern-0.2em R}}}

\def\epem       {\ensuremath{e^+e^-}\xspace}

\def\ellell     {\ensuremath{\ell^+ \ell^-}\xspace}

\def\q     {\ensuremath{q}\xspace}
\def\qbar  {\ensuremath{\overline q}\xspace}
\def\qqbar {\ensuremath{q\overline q}\xspace}
\def\u     {\ensuremath{u}\xspace}

\def\d     {\ensuremath{d}\xspace}
\def\dbar  {\ensuremath{\overline d}\xspace}

\def\s     {\ensuremath{s}\xspace}
\def\sbar  {\ensuremath{\overline s}\xspace}

\def\c     {\ensuremath{c}\xspace}

\def\b     {\ensuremath{b}\xspace}

\def\piz   {\ensuremath{\pi^0}\xspace}

\def\pip   {\ensuremath{\pi^+}\xspace}
\def\pim   {\ensuremath{\pi^-}\xspace}

\def\Kbar  {\kern 0.2em\overline{\kern -0.2em K}{}\xspace}

\def\Kz    {\ensuremath{K^0}\xspace}
\def\Kzb   {\ensuremath{\Kbar^0}\xspace}
\def\KzKzb {\ensuremath{\Kz \kern -0.16em \Kzb}\xspace}
\def\Kp    {\ensuremath{K^+}\xspace}
\def\Km    {\ensuremath{K^-}\xspace}

\def\KpKm  {\ensuremath{\Kp \kern -0.16em \Km}\xspace}

\def\Kstarz  {\ensuremath{K^{*0}}\xspace}
\def\Kstarzb {\ensuremath{\Kbar^{*0}}\xspace}

\def\Kstarm  {\ensuremath{K^{*-}}\xspace}

\def\Dbar    {\kern 0.2em\overline{\kern -0.2em D}{}\xspace}

\def\Dz      {\ensuremath{D^0}\xspace}
\def\Dzb     {\ensuremath{\Dbar^0}\xspace}
\def\DzDzb   {\ensuremath{\Dz {\kern -0.16em \Dzb}}\xspace}
\def\Dp      {\ensuremath{D^+}\xspace}
\def\Dm      {\ensuremath{D^-}\xspace}

\def\DpDm    {\ensuremath{\Dp {\kern -0.16em \Dm}}\xspace}

\def\B       {\ensuremath{B}\xspace}
\def\Bbar    {\kern 0.18em\overline{\kern -0.18em B}{}\xspace}

\def\BB      {\ensuremath{B\Bbar}\xspace} 
\def\Bz      {\ensuremath{B^0}\xspace}
\def\Bzb     {\ensuremath{\Bbar^0}\xspace}
\def\BzBzb   {\ensuremath{\Bz {\kern -0.16em \Bzb}}\xspace}
\def\Bu      {\ensuremath{B^+}\xspace}
\def\Bub     {\ensuremath{B^-}\xspace}
\def\Bp      {\ensuremath{\Bu}\xspace}
\def\Bm      {\ensuremath{\Bub}\xspace}

\def\BpBm    {\ensuremath{\Bu {\kern -0.16em \Bub}}\xspace}

\def\BorBbar    {\kern 0.18em\optbar{\kern -0.18em B}{}\xspace}
\def\DorDbar    {\kern 0.18em\optbar{\kern -0.18em D}{}\xspace}
\def\KorKbar    {\kern 0.18em\optbar{\kern -0.18em K}{}\xspace}

\def\jpsi     {\ensuremath{{J\mskip -3mu/\mskip -2mu\psi\mskip 2mu}}\xspace}
\def\psitwos  {\ensuremath{\psi{(2S)}}\xspace}

\mathchardef\Upsilon="7107
\def\Y#1S{\ensuremath{\Upsilon{(#1S)}}\xspace}

\def\FourS {\Y4S}

\mathchardef\Deltares="7101
\mathchardef\Xi="7104
\mathchardef\Lambda="7103
\mathchardef\Sigma="7106
\mathchardef\Omega="710A

\def\Deltabar{\kern 0.25em\overline{\kern -0.25em \Deltares}{}\xspace}
\def\Lbar{\kern 0.2em\overline{\kern -0.2em\Lambda\kern 0.05em}\kern-0.05em{}\xspace}
\def\Sigbar{\kern 0.2em\overline{\kern -0.2em \Sigma}{}\xspace}
\def\Xibar{\kern 0.2em\overline{\kern -0.2em \Xi}{}\xspace}
\def\Obar{\kern 0.2em\overline{\kern -0.2em \Omega}{}\xspace}
\def\Nbar{\kern 0.2em\overline{\kern -0.2em N}{}\xspace}
\def\Xb{\kern 0.2em\overline{\kern -0.2em X}{}\xspace}

\def\mes        {\mbox{$m_{\rm ES}$}\xspace}

\def\DeltaE     {\mbox{$\Delta E$}\xspace}

\newcommand{\tev}{\ensuremath{\mathrm{\,Te\kern -0.1em V}}\xspace}
\newcommand{\gev}{\ensuremath{\mathrm{\,Ge\kern -0.1em V}}\xspace}
\newcommand{\mev}{\ensuremath{\mathrm{\,Me\kern -0.1em V}}\xspace}
\newcommand{\kev}{\ensuremath{\mathrm{\,ke\kern -0.1em V}}\xspace}
\newcommand{\ev}{\ensuremath{\mathrm{\,e\kern -0.1em V}}\xspace}
\newcommand{\gevc}{\ensuremath{{\mathrm{\,Ge\kern -0.1em V\!/}c}}\xspace}
\newcommand{\mevc}{\ensuremath{{\mathrm{\,Me\kern -0.1em V\!/}c}}\xspace}
\newcommand{\gevcc}{\ensuremath{{\mathrm{\,Ge\kern -0.1em V\!/}c^2}}\xspace}
\newcommand{\mevcc}{\ensuremath{{\mathrm{\,Me\kern -0.1em V\!/}c^2}}\xspace}

\def\invfb   {\ensuremath{\mbox{\,fb}^{-1}}\xspace}

\def\mus  {\ensuremath{\rm \,\mus}\xspace}

\def\mus        {\ensuremath{\,\mu{\rm s}}\xspace}    

\def\order{{\ensuremath{\cal O}}\xspace}

\def\to                 {\ensuremath{\rightarrow}\xspace}

\def\pep2{PEP-II}

\newcommand{\dedx}{\ensuremath{\mathrm{d}\hspace{-0.1em}E/\mathrm{d}x}\xspace}

\def\gsim{{~\raise.15em\hbox{$>$}\kern-.85em
          \lower.35em\hbox{$\sim$}~}\xspace}
\def\lsim{{~\raise.15em\hbox{$<$}\kern-.85em
          \lower.35em\hbox{$\sim$}~}\xspace}

\xspace

\newcommand{\tabref}[1]{Table~\ref{tab:#1}}

\def\evtgen     {\mbox{\tt EvtGen}\xspace}

\def\geant      {\mbox{\tt GEANT}\xspace}

\def\jetset74   {\mbox{\tt Jetset \hspace{-0.5em}7.\hspace{-0.2em}4}\xspace}

\renewcommand{\eqref}[1]{Eq.~(\ref{eq:#1})}

\newcommand{\shortfigref}[1]{Fig.~\ref{fig:#1}}

\newcommand{\kppeff}       {\mbox{21.6\%}\xspace} 
\newcommand{\kkpeff}       {\mbox{17.8\%}\xspace} 
\newcommand{\kppNCand}     {\mbox{$26\,478$}\xspace}
\newcommand{\kkpNCand}     {\mbox{$7\,822$}\xspace}
\newcommand{\kppNSig}      {\mbox{22}\xspace}
\newcommand{\kkpNSig}      {\mbox{\ensuremath{-26}}\xspace}
\newcommand{\kppNSigErr}   {\mbox{43}\xspace}
\newcommand{\kkpNSigErr}   {\mbox{19}\xspace}

\newcommand{\kppNBias}     {\mbox{$3.6\pm2.4$}\xspace}
\newcommand{\kkpNBias}     {\mbox{$0.5\pm1.0$}\xspace}
\newcommand{\kppNBiasSyst} {\mbox{3.1}\xspace}
\newcommand{\kkpNBiasSyst} {\mbox{1.0}\xspace}
\newcommand{\kppNPDFSyst}  {\mbox{10.0}\xspace}
\newcommand{\kkpNPDFSyst}  {\mbox{3.5}\xspace}
\newcommand{\kppEffSyst}   {\mbox{13.0\%}\xspace}
\newcommand{\kkpEffSyst}   {\mbox{13.5\%}\xspace}
\newcommand{\kppVetoSyst}  {\mbox{$^{+0\%}_{-18\%}$}\xspace}
\newcommand{\kkpVetoSyst}  {\mbox{$^{+25\%}_{-0\%}$}\xspace}
\newcommand{\kppBF}        {( 1.8 \pm 4.3 \pm 0.9 ) \times 10^{-7}}
\newcommand{\kkpBF}        {( -3.2 \pm 2.3 \,^{+1.0}_{-0.6} ) \times 10^{-7}}
\newcommand{\kppSens}      {\mbox{$7.4\times10^{-7}$}\xspace} 
\newcommand{\kkpSens}      {\mbox{$4.2\times10^{-7}$}\xspace} 
\newcommand{\kppUL}        {\mbox{$9.5\times10^{-7}$}\xspace} 
\newcommand{\kkpUL}        {\mbox{$1.6\times10^{-7}$}\xspace} 

\newcommand{\onreslumi}  {\mbox{426\invfb}}
\newcommand{\offreslumi} {\mbox{44\invfb}}
\newcommand{\bbpairs}    {\mbox{$(467\pm5)\times10^{6}$}}

\newcommand{\Lzero}      {\mbox{$L_0$}}
\newcommand{\Ltwo}       {\mbox{$L_2$}}

\newcommand{\Kpppos}             {\mbox{$\Kp   \pim  \pip$}}

\newcommand{\BtoKpppos}          {\mbox{$\Bp \to \Kpppos$}}

\newcommand{\pppneg}             {\mbox{$\pim  \pip  \pim$}}
\newcommand{\Kppneg}             {\mbox{$\Km   \pip  \pim$}}

\newcommand{\KKKneg}             {\mbox{$\Km   \Kp   \Km $}}
\newcommand{\KPPSMSneg}          {\mbox{$\Kp   \pim  \pim$}}
\newcommand{\KKPSMSneg}          {\mbox{$\Km   \Km   \pip$}}

\newcommand{\Btopppneg}          {\mbox{$\Bm \to \pppneg$}}
\newcommand{\BtoKppneg}          {\mbox{$\Bm \to \Kppneg$}}

\newcommand{\BtoKKKneg}          {\mbox{$\Bm \to \KKKneg$}}
\newcommand{\BtoKPPSMSneg}       {\mbox{$\Bm \to \KPPSMSneg$}}
\newcommand{\BtoKKPSMSneg}       {\mbox{$\Bm \to \KKPSMSneg$}}

\newcommand{\KstarI}             {\mbox{$\Kstarz(892)$}}

\newcommand{\KstarII}            {\mbox{$\Kstarz_{0}(1430)$}}

\newcommand{\Dzpim}              {\mbox{$\Dz \pim$}}
\newcommand{\BmtoDzpim}          {\mbox{$\Bm \to \Dzpim$}}

\newcommand{\DztoKpi}           {\mbox{$\Dz \to \Km\pip$}}
\newcommand{\DztoKK}            {\mbox{$\Dz \to \Km\Kp$}}
\newcommand{\DztoKpipiz}        {\mbox{$\Dz \to \Km\pip\piz$}}

\newcommand{\JPsitoll}           {\mbox{$\jpsi \to \ellell$}}

\newcommand{\Psitoll}            {\mbox{$\psitwos \to \ellell$}}

\newcommand{\BztoKpipiz}         {\mbox{$\Bz \to \Kp\pim\piz$}}

\newcommand{\BztoKppim}          {\mbox{$\Bz \to \Kp \pim$}}

\def\nnout    {\ensuremath{{\rm NN}_{\rm out}}}

\newcommand{\BABARPubYear}    {08}
\newcommand{\BABARPubNumber}  {033}
\newcommand{\SLACPubNumber} {13356}
\newcommand{\LANLNumber} {0808.0900}

\def\figurebox#1#2#3{%
    \def\arg{#3}%
    \ifx\arg\empty
    {\hfill\vbox{\hsize#2\hrule\hbox to #2{\vrule\hfill\vbox to #1{\hsize#2\vfill}\vrule}\hrule}\hfill}%
    \else
    {\hfill\epsfbox{#3}\hfill}%
    \fi}

\begin{document}


\begin{flushleft}
arXiv:\LANLNumber\ [hep-ex] \\
SLAC-PUB-\SLACPubNumber \\
\babar-PUB-\BABARPubYear/\BABARPubNumber
\end{flushleft}

\title{
  {
    \large \bf \boldmath 
    Search for the highly suppressed decays \BtoKPPSMSneg\ and \BtoKKPSMSneg
  }
}

%
\author{B.~Aubert}
\author{M.~Bona}
\author{Y.~Karyotakis}
\author{J.~P.~Lees}
\author{V.~Poireau}
\author{E.~Prencipe}
\author{X.~Prudent}
\author{V.~Tisserand}
\affiliation{Laboratoire de Physique des Particules, IN2P3/CNRS et Universit\'e de Savoie, F-74941 Annecy-Le-Vieux, France }
\author{J.~Garra~Tico}
\author{E.~Grauges}
\affiliation{Universitat de Barcelona, Facultat de Fisica, Departament ECM, E-08028 Barcelona, Spain }
% ...
\collaboration{The \babar\ Collaboration}
\noaffiliation

\date{
  \today
}

\begin{abstract} 

We report a search for the decays \BtoKPPSMSneg\ and \BtoKKPSMSneg, which
are highly suppressed in the standard model.
Using a sample of \bbpairs\ \BB\ pairs collected with the \babar\ detector,
we do not see any evidence of these decays and determine 90\% confidence level upper limits of
${\cal B}(\BtoKPPSMSneg) < \kppUL$ and
${\cal B}(\BtoKKPSMSneg) < \kkpUL$ on the corresponding branching fractions, including systematic uncertainties.

\end{abstract}

\maketitle

\begin{figure}[!htb]    
\includegraphics[width=\columnwidth]{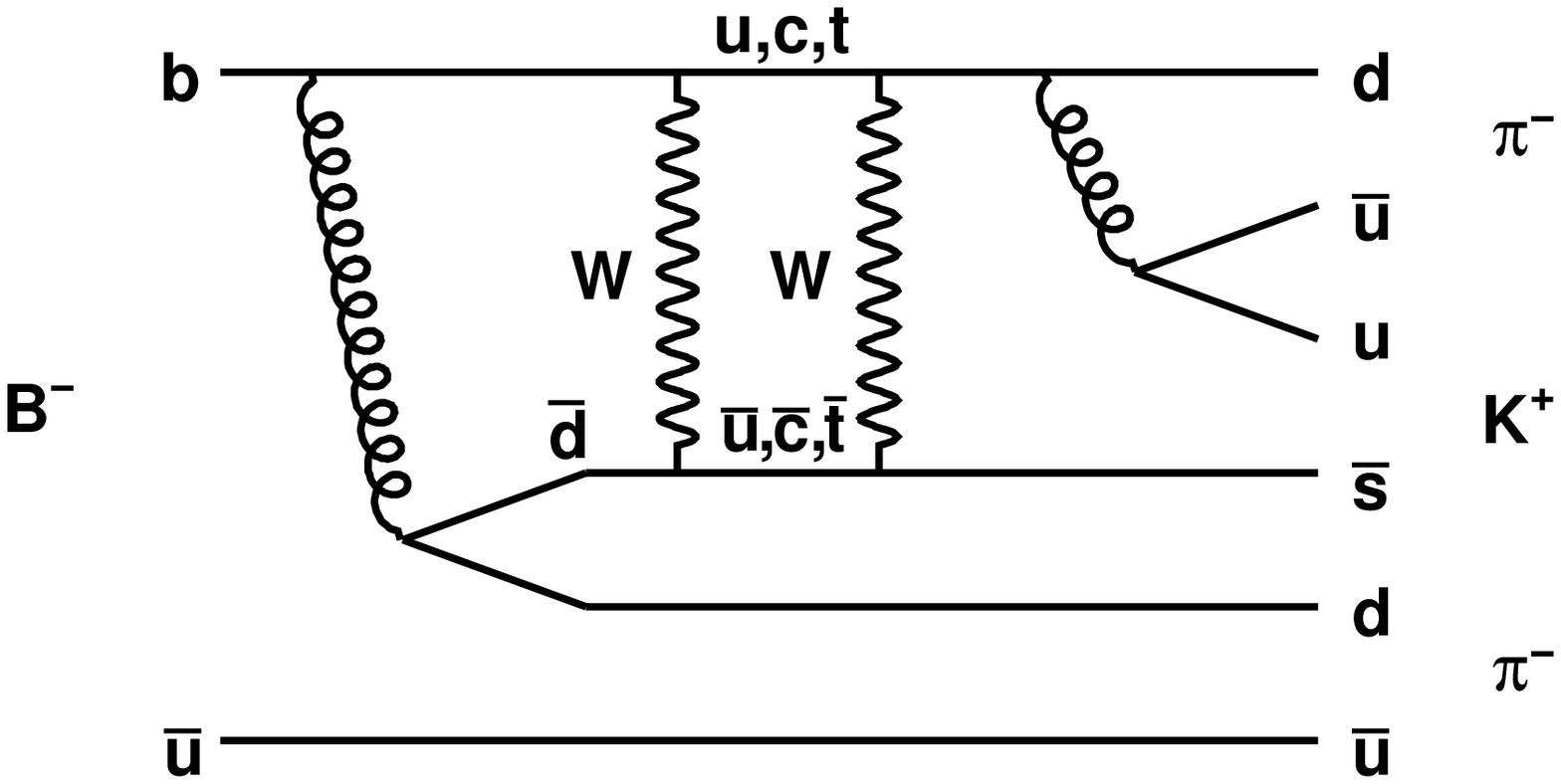}
\includegraphics[width=\columnwidth]{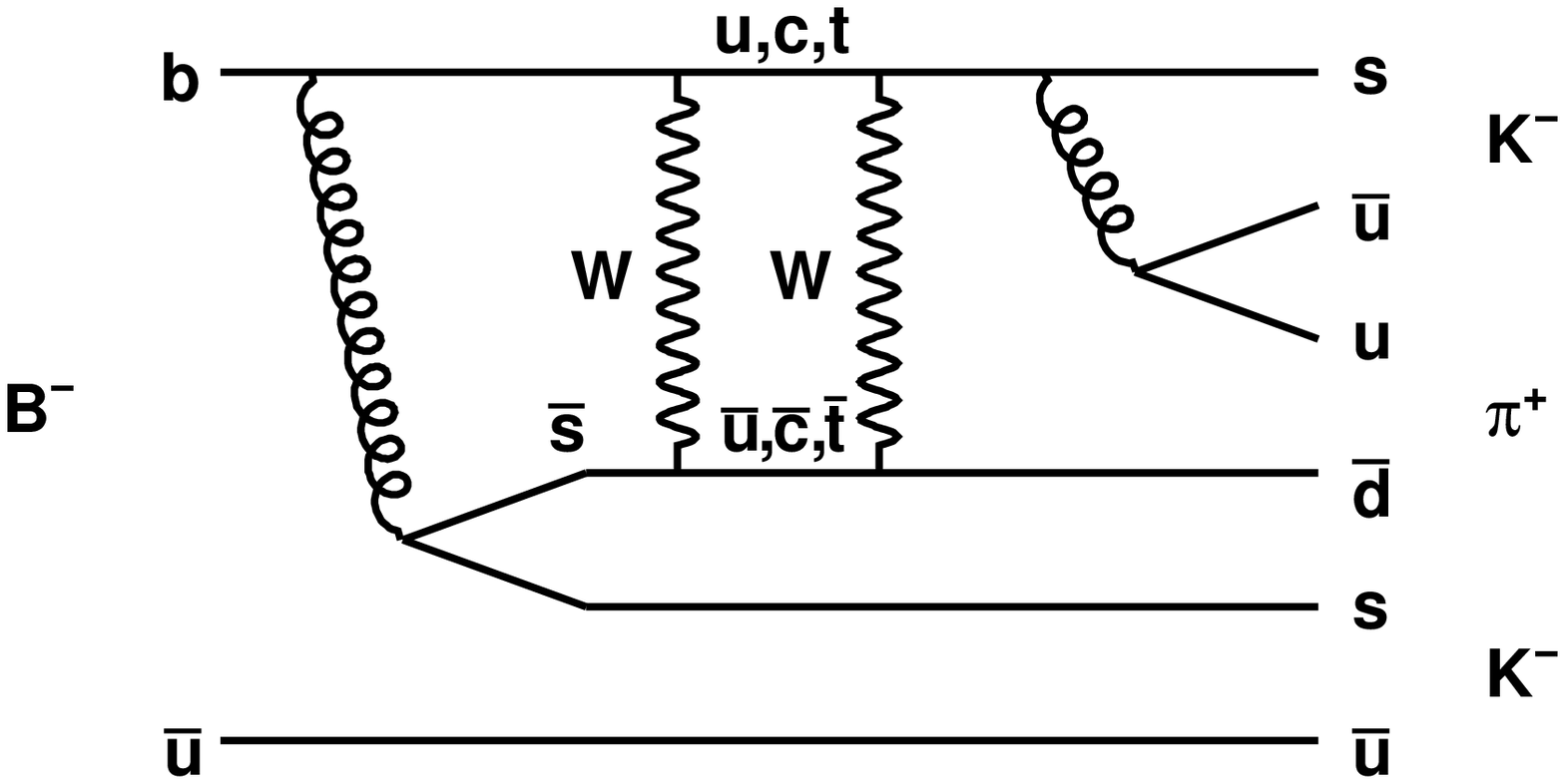}
\caption{
  Example standard model decay diagrams for the decays
  \BtoKPPSMSneg\ and \BtoKKPSMSneg, respectively.
}
\label{fig:feynman}
\end{figure}

The decays \BtoKPPSMSneg\ and \BtoKKPSMSneg\ proceed via $\b\to\d\d\sbar$
and $\b\to\s\s\dbar$ quark transitions, respectively.
These are highly suppressed in the standard model (SM).  
Compared with the penguin (loop) transitions $\b\to\q\qbar\d$, and
$\b\to\q\qbar\s$ they are additionally suppressed by the small
Cabibbo-Kobayashi-Maskawa matrix~\cite{Cabibbo:1963yz,Kobayashi:1973fv}
element factor $\left| V_{td}V_{ts}^* \right| \simeq 3 \times 10^{-4}$,
leading to predicted branching fractions of only $\order(10^{-14})$ and
$\order(10^{-11})$, respectively~\cite{Huitu:1998vn,Fajfer:2006av}. 
Example SM decay diagrams can be seen in \shortfigref{feynman}.

These branching fractions can be significantly enhanced in SM extensions
such as the minimal supersymmetric standard model (MSSM) with or
without conserved $R$ parity, or in models containing extra $U(1)$ gauge
bosons.  For example, the branching fraction for the $\b\to\d\d\sbar$ transition
$\Bm\to\pim\Kstarz$ can be enhanced from about $10^{-16}$ in the SM to about
$10^{-6}$ in the presence of an extra $Z^\prime$ boson~\cite{Fajfer:2006av}.
The branching fraction for the $\b\to\s\s\dbar$ decay $\Bm\to\Km\Kstarzb$
can be enhanced from about $7 \times 10^{-14}$ in the SM to about $6 \times
10^{-9}$ in the MSSM~\cite{Fajfer:2000ny}.

Observations of the decays \BtoKPPSMSneg\ and \BtoKKPSMSneg\ would be clear
experimental signals for the $\b\to\d\d\sbar$ and $\b\to\s\s\dbar$ quark
transitions~\cite{Chun:2003rg,Browder:2008em}.
These decay modes have been previously searched
for~\cite{Bergfeld:1996dd,Abbiendi:1999st,Aubert:2003xz,Garmash:2003er},
and the most restrictive 90\% confidence level experimental upper limits
${\cal B}(\BtoKPPSMSneg) < 1.8 \times 10^{-6}$ and
${\cal B}(\BtoKKPSMSneg) < 1.3 \times 10^{-6}$~\cite{Aubert:2003xz}
were obtained from an analysis of 81.8 \invfb\ of \babar\ data.
Upper limits on $\b\to\s\s\dbar$ transitions have also been set using the
decays
$\Bm\to\Kstarm\Km\pip$~\cite{Aubert:2006aw}, 
$\Bzb\to\Kstarzb\Km\pip$~\cite{Aubert:2007fm}, and 
$\Bzb\to\Kstarzb\Kstarzb$~\cite{Aubert:2007xc}.

We report the results of a search for the decays \BtoKPPSMSneg\ and
\BtoKKPSMSneg.
Inclusion of the charge conjugate modes is implied throughout this paper.
The data used in this analysis, collected
at the \pep2\ asymmetric energy \epem collider~\cite{pep2}, 
consist of an integrated luminosity of \onreslumi\ 
recorded at the \FourS\ resonance. 
In addition, \offreslumi\ of data were collected 40\,\mev\ below 
the resonance and are used for background characterization.
These samples are referred to as on-resonance and off-resonance data,
respectively.
The on-resonance data sample contains \bbpairs\ \BB\ pairs.
Beyond the larger sample size, we utilize improved analysis
techniques for background rejection and signal identification compared with
our previous study~\cite{Aubert:2003xz}.

The \babar\ detector is described in detail elsewhere~\cite{Aubert:2001tu}.
Charged particles are detected and their momenta measured with a five-layer
silicon vertex tracker and a 40-layer drift chamber (DCH) inside a
1.5\,T solenoidal magnet.
Surrounding the DCH is a detector of internally reflected Cherenkov
radiation.
Energy deposited by electrons and photons is measured by a CsI(Tl) crystal
electromagnetic calorimeter.

We select \BtoKPPSMSneg\ candidates by combining a charged kaon candidate
with two charged pion candidates, each of which has charge opposite to the
kaon.  Similarly, \BtoKKPSMSneg\ candidates are selected by combining two
charged kaon candidates with a charged pion candidate.
Each track is required to have a minimum transverse momentum of 50\,\mevc, 
and to be consistent with having originated from the interaction region. 
Identification of charged pions and kaons is accomplished using energy loss
(\dedx) information from the silicon vertex tracker and DCH, and the
Cherenkov angle and number of photons measured in the detector of
internally reflected Cherenkov radiation.
The efficiency for kaon selection is approximately 80\% including
geometrical acceptance, while the probability of misidentification of pions
as kaons is below 5\%.
The corresponding efficiency and kaon misindentification rate for the pion
selection criteria are 95\% and less than 10\%, respectively.
We require all charged particle candidates to be inconsistent with the
electron hypothesis, 
based on a cut-based selection algorithm that uses 
information from \dedx, shower shapes in the electromagnetic calorimeter, 
and the ratio of the shower energy and track momentum.

To avoid a potentially large source of background arising from \B\ decays
mediated by the favored $b \to c$ transition, we veto \B\ candidates for
which pairs of daughter tracks have invariant mass combinations in the
ranges $1.76 < m_{K\pi} < 1.94 \gevcc$, $2.85 < m_{K\pi} < 3.25 \gevcc$,
and $3.65 < m_{K\pi} < 3.75 \gevcc$.
These remove events containing the decays \DztoKpi, \JPsitoll, and 
\Psitoll, respectively, where the leptons in the \jpsi\ and \psitwos\ decays
are misidentified as pions or kaons.

Continuum $\epem\to\qqbar \ (\q=\u,\d,\s,\c)$ events are the dominant
background. 
To discriminate this type of event from signal,
we use a neural network that combines five variables.
The first of these is the ratio of \Ltwo\ to \Lzero, with
$L_j = \sum_i p^{\star}_i \left|\cos\theta^{\star}_i\right|^j$,
where $p^{\star}_i$ is the particle momentum, $\theta^{\star}_i$ is the
angle between the particle and the thrust axis determined from the
\B\ candidate decay products, the sum is over all tracks and neutral
clusters not associated with the \B\ candidate, and all quantities are
calculated in the \epem\ center-of-mass (CM) frame.
The other four variables are the absolute value of the cosine of the angle
between the \B direction and the beam axis;
the magnitude of the cosine of the angle between the \B\ thrust axis 
and the beam axis;
the product of the \B\ candidate's charge and the output of a
multivariate algorithm that identifies the flavor of the recoiling
\B\ meson~\cite{Aubert:2004zt};
and the proper time difference between the decays of the two \B\ mesons
divided by its uncertainty.
The angles with respect to the beam axis are calculated in the CM frame.
The neural network output \nnout\ is distributed such that it peaks around
0 for continuum background and around 1 for signal.  We require $\nnout >
0.5$ ($\nnout > 0.4$) for \BtoKPPSMSneg\ (\BtoKKPSMSneg) candidates.
These requirements retain approximately 90\% of the signal, while rejecting
approximately 80\% of the continuum background.

In addition to the neural network output,
we distinguish signal from background events using two kinematic variables: 
the difference \DeltaE\ between the CM energy
of the \B\ candidate and $\sqrt{s}/2$, 
and the beam-energy substituted mass
$\mes=\sqrt{s/4-{\bf p}^{\star2}_\B}$,
where $\sqrt{s}$ is the total CM energy
and ${\bf p}^{\star}_\B$ is the momentum of the candidate \B\ meson in the CM frame.
The \DeltaE\ distribution peaks near zero with a resolution of around
$19\mev$, while the \mes\ distribution for signal events peaks near the
\B\ mass with a resolution of around $2.4\mevcc$.
We select signal candidates that satisfy
$5.260<\mes<5.286\gevcc$ and $|\DeltaE|<0.100\gev$.
This region includes a sufficiently large range of \mes\ below the signal
peak to determine properties of the continuum distribution.

The efficiency for signal events to pass the selection criteria is
\kppeff\ (\kkpeff) for \BtoKPPSMSneg\ (\BtoKKPSMSneg),
determined with a Monte Carlo (MC) simulation in which the decays are
generated uniformly in three-body phase space.
The \babar\ detector Monte Carlo simulation is based on
\geant4~\cite{Agostinelli:2002hh} and \evtgen~\cite{Lange:2001uf}.
We find that 8.2\% (5.1\%) of \BtoKPPSMSneg\ (\BtoKKPSMSneg) selected
events contain more than one candidate, in which case we choose the one with
the highest probability that the three tracks originate from a common vertex. 

We study possible residual backgrounds from \BB\ events using MC event samples.
Backgrounds arise from decays with topologies similar to the signal but
with some misreconstruction.  Such effects include kaon/pion
misidentification, the loss of a soft neutral particle, and the association
of a particle from the decay of the other \B\ in the event with the signal
candidate or vice versa.
We find that the backgrounds can be conveniently divided into five
categories for both the \KPPSMSneg\ and \KKPSMSneg\ channels,
each of which is dominated by one or two particular decays but also
includes other decay modes that result in similar \mes\ and \DeltaE\ shapes. 
\tabref{bb} provides details of the composition of the background categories.

\begin{table*}[!htb]
\caption
{
Summary of the \B\ background categories, giving the dominant decay mode,
numbers of expected and observed events and the character of the \mes\ and
\DeltaE\ distributions.
``Peaking'' indicates that the shape is similar to that of the signal.
``Broad peak,'' ``left peak,'' and ``right peak'' differ from the signal in
being wider or shifted to lower or higher values, respectively.
The number of expected and observed events are also given for the continuum
background.
}
\label{tab:bb}
\begin{center}
\resizebox{\textwidth}{!}{
\begin{tabular}{lcccccc}
\hline
\multicolumn{7}{l}{\BtoKPPSMSneg} \\
\hline
Category                  & 1            & 2           & 3              & 4              & 5              & Continuum     \\
\hline                                                                                                                    
Dominant mode(s)          & \BmtoDzpim;  & \Btopppneg  & \BtoKppneg \&  & \BztoKppim     & Generic \BB    & $\cdots$      \\
                          & \DztoKK      &             & \BztoKpipiz    &                &                &               \\
Number of expected events & $80\pm3$     & $57\pm4$    & $472\pm24$     & $43\pm1$       & $917\pm19$     & $25552\pm495$ \\
Number of observed events & $61\pm70$    & $-153\pm94$ & $1116\pm347$   & $-26\pm152$    & $197\pm273$    & $25261\pm198$ \\
\mes\ Structure           & Peaking      & Peaking     & Broad peak     & Broad peak     & Continuum-like & $\cdots$      \\
\DeltaE\ Structure        & Left peak    & Right peak  & Broad peak     & Right peak     & Continuum-like & $\cdots$      \\
\hline
\multicolumn{7}{l}{\BtoKKPSMSneg} \\
\hline
Category                  & 1            & 2           & 3              & 4              & 5              & Continuum     \\
\hline                                                                                                                    
Dominant mode(s)          & \BtoKKKneg   & \BtoKppneg  & \BmtoDzpim;    & Generic \BpBm  & Generic \BzBzb & $\cdots$      \\
                          &              &             & \DztoKpipiz    &                &                &               \\
Number of expected events & $190\pm9$    & $198\pm9$   & $61\pm4$       & $312\pm11$     & $173\pm8$      & $6088\pm241$  \\
Number of observed events & $213\pm41$   & $240\pm37$  & $-34\pm55$     & $380\pm117$    & $95\pm107$     & $6953\pm100$  \\
\mes\ Structure           & Peaking      & Peaking     & Broad peak     & Broad peak     & Continuum-like & $\cdots$      \\
\DeltaE\ Structure        & Left peak    & Right peak  & Left peak      & Continuum-like & Continuum-like & $\cdots$      \\
\hline
\end{tabular}
}
\end{center}
\end{table*}

In order to obtain the \BtoKPPSMSneg\ and \BtoKKPSMSneg\ signal yields,
we perform unbinned extended maximum likelihood fits to the
candidate events using three variables: \mes, \DeltaE, and \nnout.
For each event hypothesis $j$ 
(signal, continuum background, or one of the five \BB\ background categories),
we define a probability density function (PDF)
\begin{equation}
  \label{PDF-exp}
  {\cal P}^i_j \equiv
  {\cal P}_j(\mes^i,\DeltaE^i)\cdot {\cal P}_j(\nnout^i) \, ,
\end{equation}
where $i$ denotes the event index. 
For the signal, continuum background, and the \BB\ background categories
with small correlations between \mes\ and \DeltaE, the PDF is further
factorized
\begin{equation}
  \label{PDF-mesde}
  {\cal P}_j(\mes^i,\DeltaE^i) = {\cal P}_j(\mes^i) \cdot {\cal P}_j(\DeltaE^i) \, .
\end{equation}
The extended likelihood function is
\begin{equation}
  \label{eq:extML-Eq}
  {\cal L} = 
  \exp\left(-\sum_{k}n_k\right)
  \prod_{i}\left[ \sum_{j}n_j{\cal P}^i_j \right],
\end{equation}
where $n_{j}$ ($n_{k}$) is the yield belonging to the event hypothesis $j$ ($k$).

The signal \mes\ and \DeltaE\ shapes are parametrized with the sum
of a Gaussian and a Crystal Ball
function~\cite{Oreglia:1980cs,Gaiser:1982yw,Skwarnicki:1986xj} and the sum of
two Gaussian functions, respectively.
We determine the shape parameters by taking the values obtained from signal MC
and correcting for differences between data and MC seen in a control
sample of \BmtoDzpim\ with \DztoKpi\ decays.
The continuum background \mes\ shape is described by the function
$x\sqrt{1-x^2}\exp\left[-\xi(1-x^2)\right]$, with $x\equiv 2\mes/\sqrt{s}$
and $\xi$ a free parameter~\cite{Albrecht:1990am},
while the continuum \DeltaE\ shape is modeled with a linear function.
We describe the \mes\ and \DeltaE\ shapes of each  \BB\ background category
using either independent 1D histograms or a 2D histogram determined from MC
samples.  The decision to use 1D or 2D histograms is made based on the
magnitude of the correlations between these variables for each category and
the effect on the signal yield of neglecting such correlations, discussed
below.
The PDFs for categories 1, 2, and 3, for both \BtoKPPSMSneg\ and
\BtoKKPSMSneg, are modeled using 2D histograms.
We use 1D histograms to describe all \nnout\ distributions. 
These histograms are obtained from MC samples for the signal and
\BB\ background categories, and from a combination of on-resonance data, in
a continuum-dominated sideband of \mes\ and \DeltaE, and off-resonance
data for the continuum background.

The nine free parameters in our fits are the yields of the signal,
continuum and all five \BB\ background categories, the $\xi$ parameter of
the continuum \mes\ shape, and the slope of the continuum \DeltaE\ shape.

We test the fitting procedure by applying it to ensembles
of simulated experiments where events are generated from the PDF shapes
described above for all seven categories of events. We repeat the
exercise with \qqbar\ events generated from the PDF while signal 
events are randomly extracted from the MC samples.  
The \BB\ background events are either generated from PDF shapes or drawn
from MC samples.
In all cases, these tests confirm that our fit performs as expected, with
very small biases on the fitted signal yields, for which we correct the
measured yields and include systematic uncertainties.

\begin{figure*}[!htb]    
\includegraphics[width=0.32\textwidth]{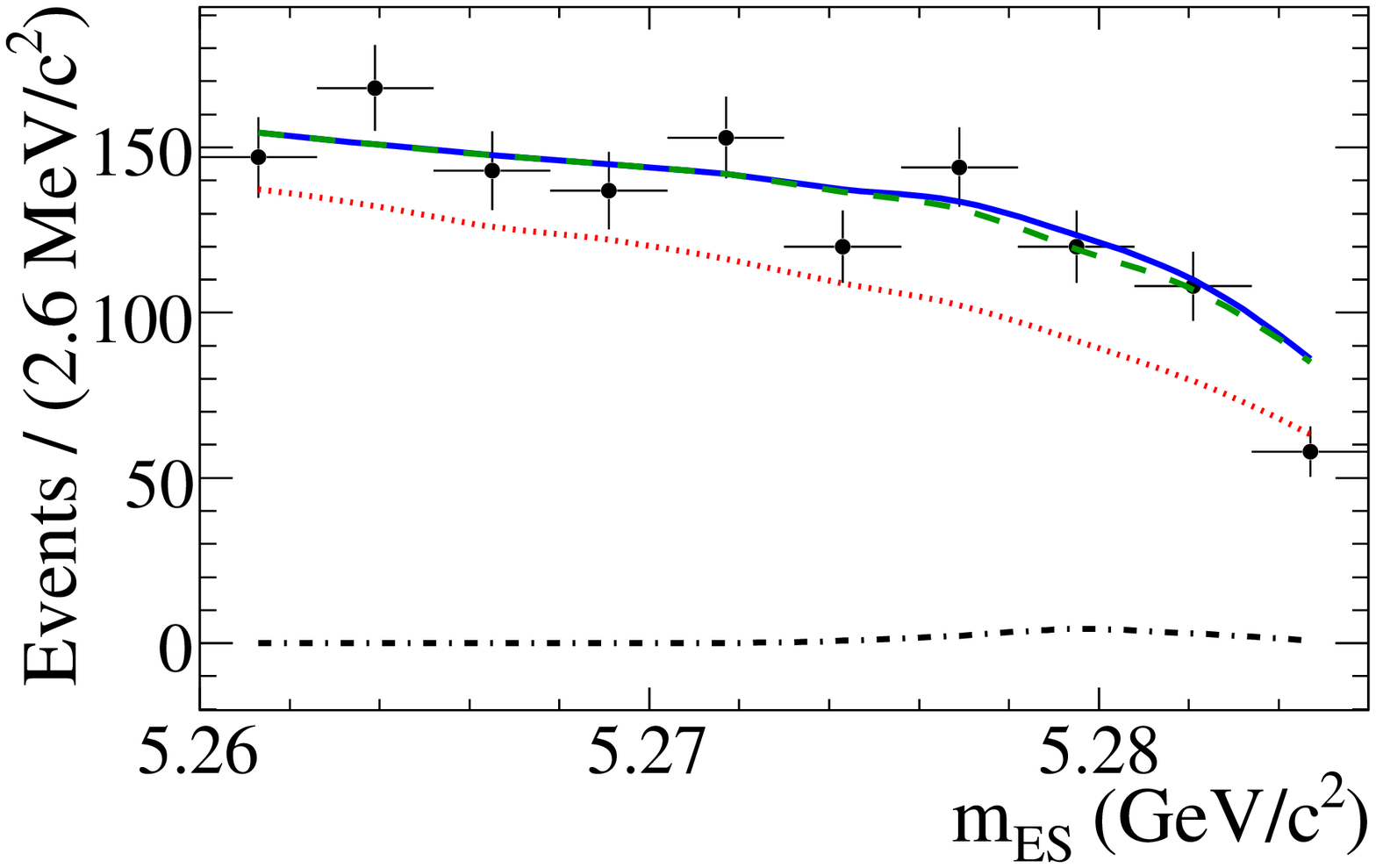}
\includegraphics[width=0.32\textwidth]{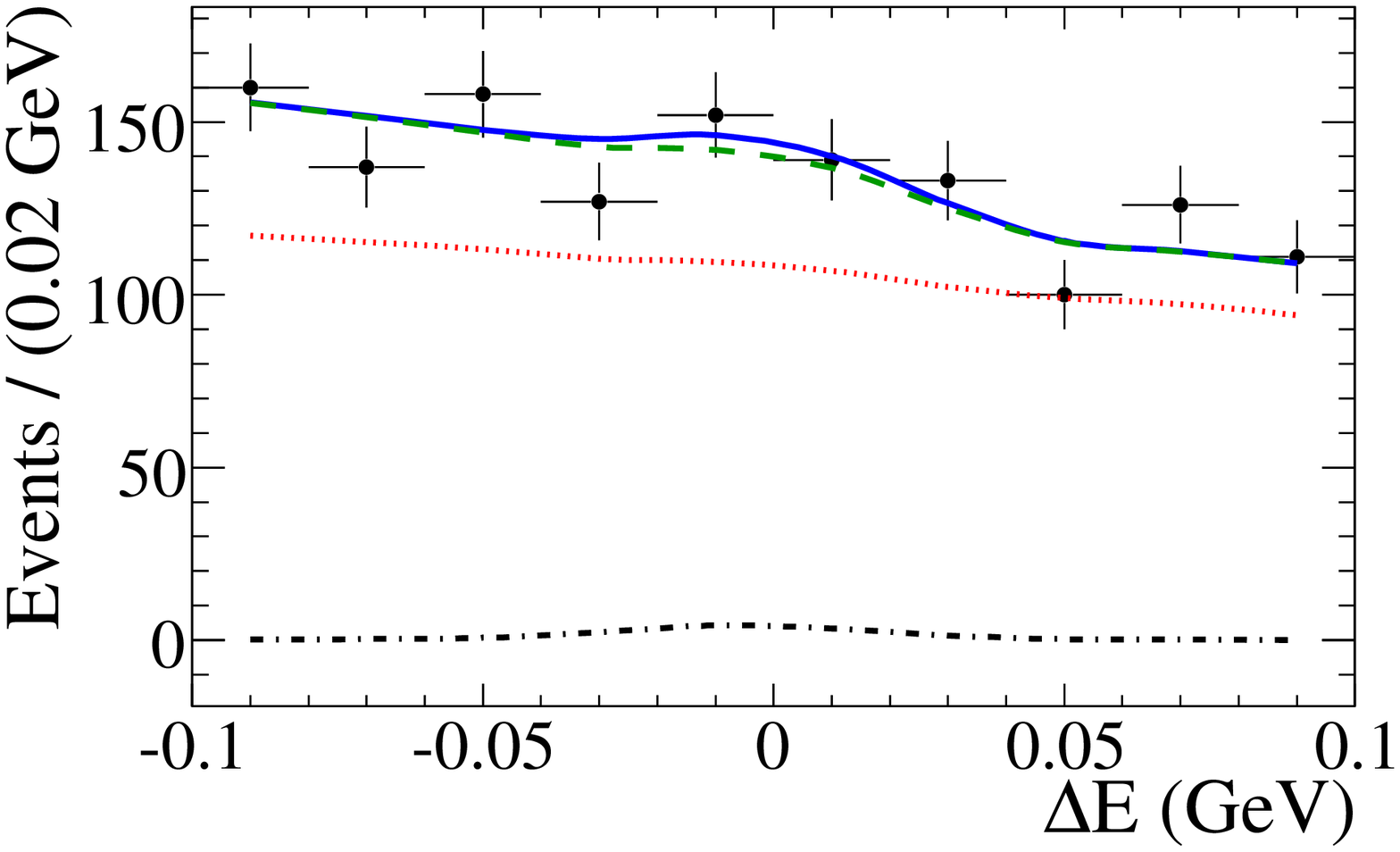}
\includegraphics[width=0.32\textwidth]{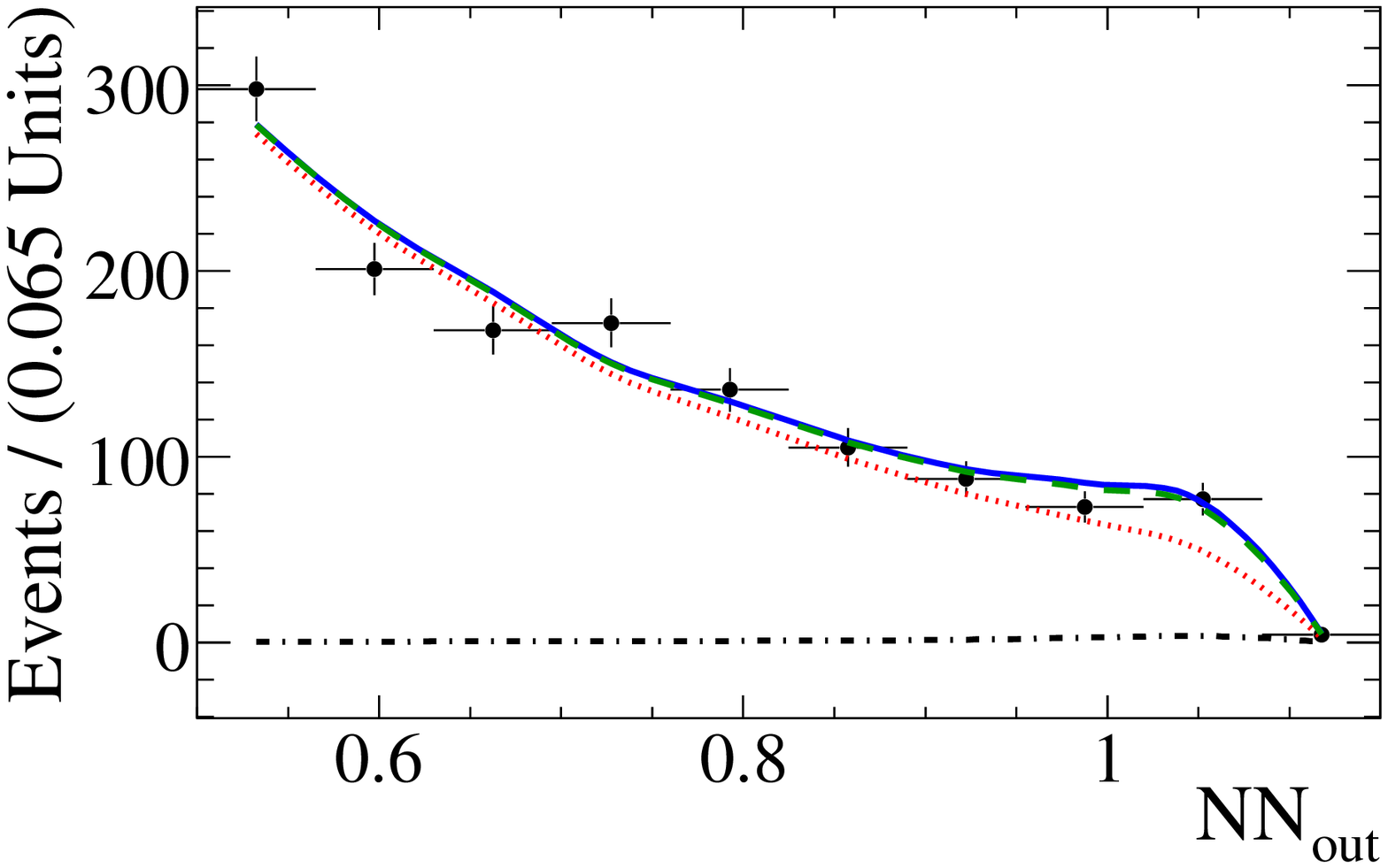}\\
\includegraphics[width=0.32\textwidth]{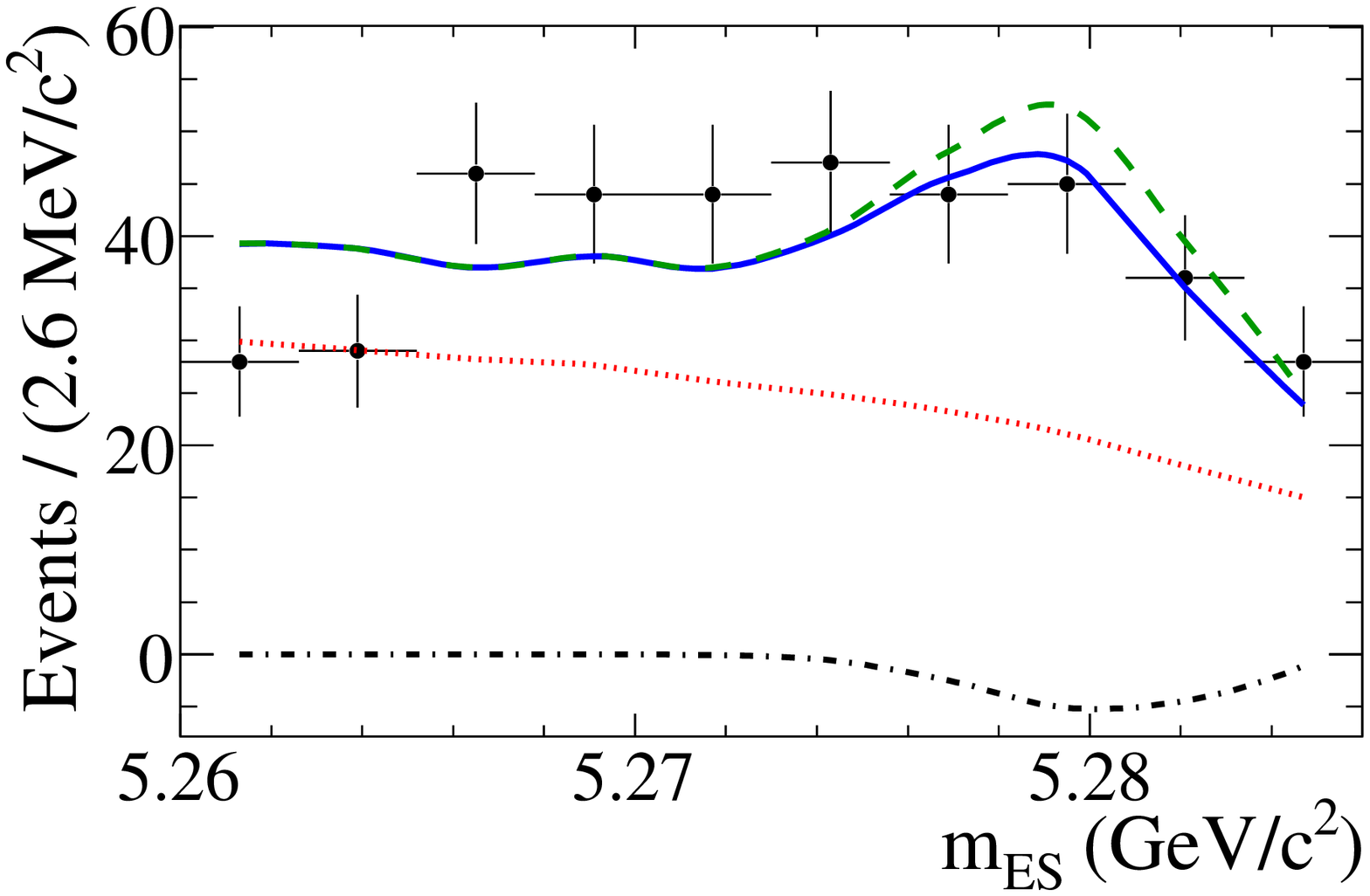}
\includegraphics[width=0.32\textwidth]{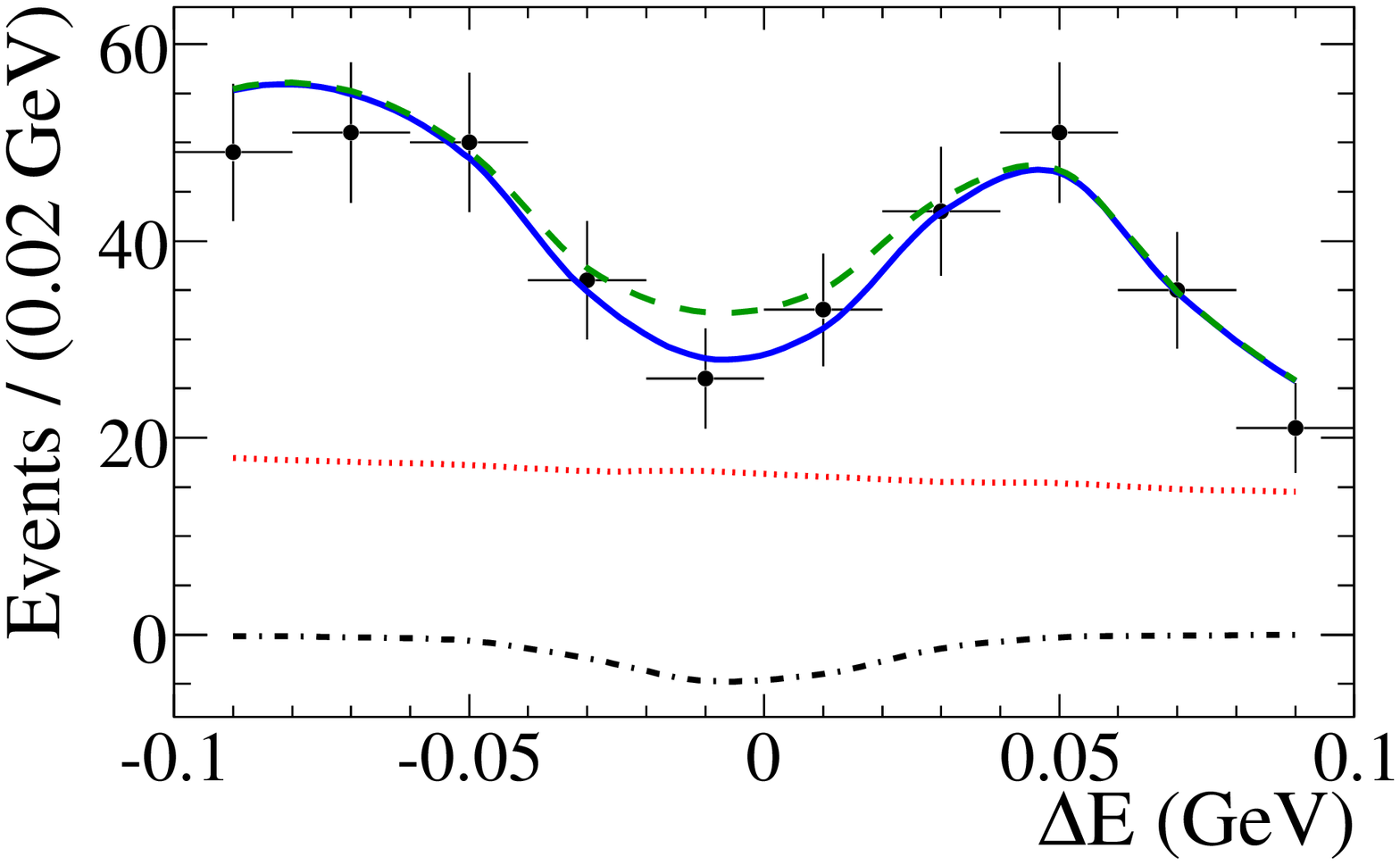}
\includegraphics[width=0.32\textwidth]{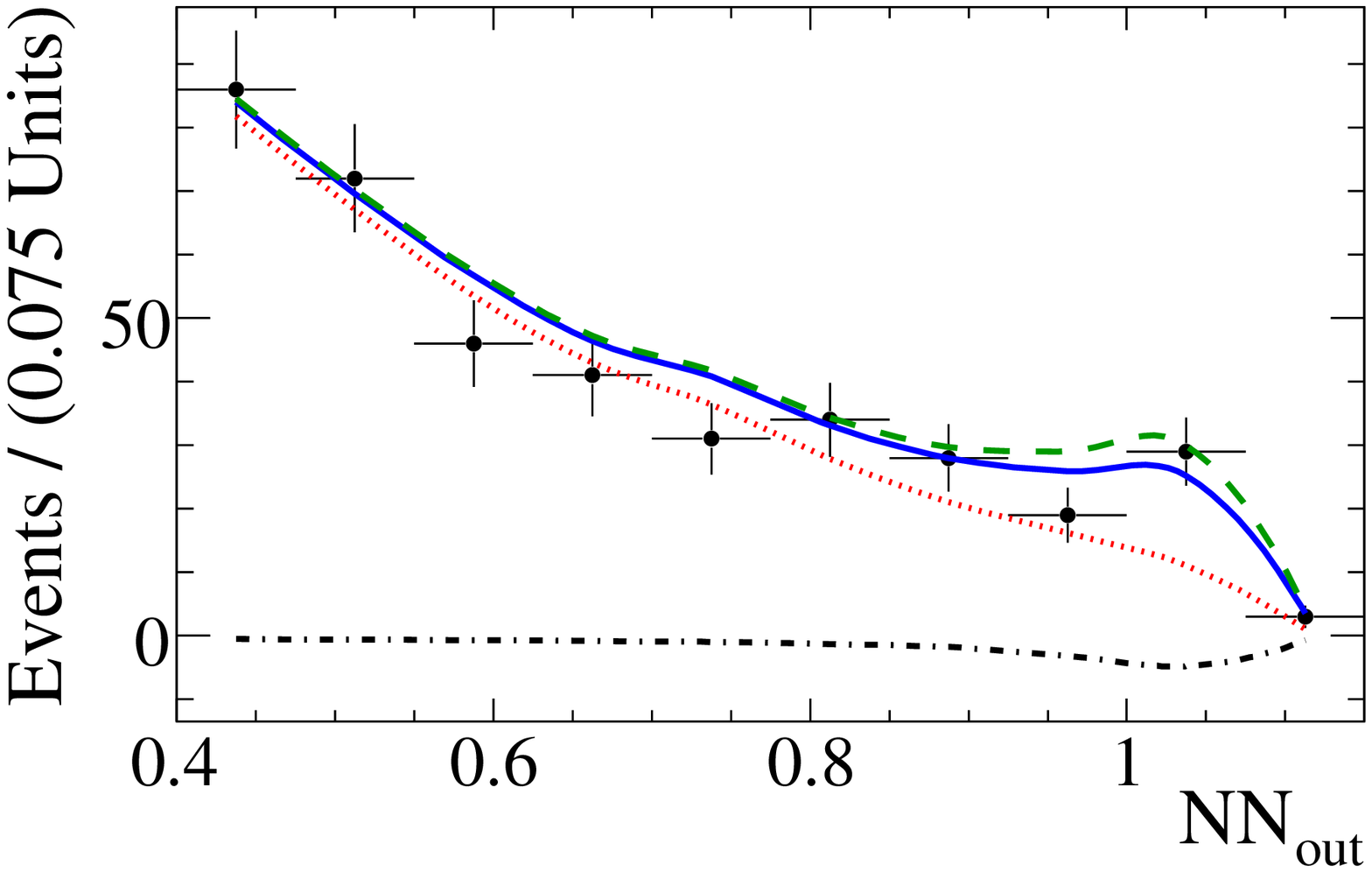}
\caption{
  Projections of the selected events with the fit results overlaid.
  The top (bottom) set of plots are for \BtoKPPSMSneg\ (\BtoKKPSMSneg).
  From left to right the plots show the projections onto \mes, \DeltaE, 
  and the output of the neural network.
  The black points are the data, the solid blue curve is the total fit, the
  dotted red curve is the continuum background, the dashed green curve is the
  total background, and the dash-dotted black curve at the bottom is the
  signal.
  The continuum component has been suppressed in these plots by applying an
  additional requirement on the ratio of the signal likelihood to the sum of
  the signal and continuum likelihoods, calculated without use of the plotted
  variable.  The value of the requirement for each plot is chosen to reject
  about 95\% of the continuum background while retaining about 55\% of the
  signal.
}
\label{fig:fit-results}
\end{figure*}

We apply the fit described above to the \kppNCand\ \BtoKPPSMSneg\ and
\kkpNCand\ \BtoKKPSMSneg\ candidate events selected from the data recorded at
the \FourS\ resonance.
We find $\kppNSig \pm \kppNSigErr$ and $\kkpNSig \pm \kkpNSigErr$ signal
events, respectively, (statistical uncertainties only).
The yields of continuum and all \BB\ background categories (shown in
\tabref{bb}) are generally consistent with expectations.
The yields of \BB\ background categories 3 and 5 from the fit to
\BtoKPPSMSneg\ candidates do not show perfect agreement; however, the sum of
their yields is consistent with the expectation and, owing to the strong
negative correlation between the yields of these categories, the discrepancy
with the expectation is not significant.  Such behavior was seen
in the fit validations and has been shown not to effect the signal yield.
The results of the fits are shown in \shortfigref{fit-results}.

We determine the branching fractions for \BtoKPPSMSneg\ and
\BtoKKPSMSneg\ by applying corrections for the small biases evaluated in
the MC studies (\kppNBias\ and \kkpNBias\ events, respectively) and then
dividing by the selection efficiencies and the total number of \BB\ pairs
in the data sample.
We assume equal decay rates of $\FourS \to \BpBm$ and $\BzBzb$.
Systematic uncertainties on the fitted yields arise from uncertainties in
the PDF shapes (\kppNPDFSyst\ and \kkpNPDFSyst\ events, respectively)
including possible data/MC differences in the signal PDF shapes
studied using the \BmtoDzpim\ control samples discussed above.
We estimate the uncertainty on the fit bias (\kppNBiasSyst\ and
\kkpNBiasSyst\ events, respectively) to be half the value of the
correction combined in quadrature with the precision with which the bias
is known.
Uncertainties on the efficiency arise from possible data/MC differences for
tracking (1.2\%) and particle identification (4.2\%).
We consider two sources of uncertainty related to the Dalitz plot
distributions of the signal decays.  The first is related to the variation
of the efficiency over the parts of the Dalitz plots that are included in
the analysis: from MC studies, the uncertainties are found to be
\kppEffSyst\ for \BtoKPPSMSneg\ and \kkpEffSyst\ for \BtoKKPSMSneg.
The second is due to the correction for the vetoed parts of the Dalitz
plots, which we estimate for various signal decay distributions.
In addition to the nominal phase-space distribution, we consider decays
dominated by the intermediate states \KstarI\ or \KstarII\ (modeled using
the LASS~\cite{Aston:1987ir} shape, as implemented in our Dalitz plot
analysis of \BtoKpppos~\cite{Aubert:2008bj}).  We mimic a possible
enhancement at low $\pim\pim$ or $\Km\Km$ invariant mass by employing an
{\it ad hoc} doubly charged scalar resonance with mass 1500 \mevcc\ and
width 300 \mevcc.
The efficiency of the veto requirement is larger than that for the
phase-space MC in all alternative models, so we assign asymmetric
systematic errors of \kppVetoSyst\ for \BtoKPPSMSneg\ and \kkpVetoSyst\ for
\BtoKKPSMSneg.  The uncertainty on the number of \BB\ pairs is $1.1\%$.
Including all systematic uncertainties, we obtain the following results for
the branching fractions:
${\cal B}(\BtoKPPSMSneg) = \kppBF$ and 
${\cal B}(\BtoKKPSMSneg) = \kkpBF$,
where the first uncertainties are statistical and the second are
systematic.

We have also calculated the branching fractions using event-by-event
efficiencies applied to signal weights obtained from the fit
result~\cite{Pivk:2004ty,Aubert:2007xb}.
We obtain results consistent with our main results within the efficiency
variation systematic uncertainty.
We have also checked that removing each of the discriminating variables from
the fit, in turn, gives consistent results.

\begin{figure*}[!htb]    
\includegraphics[width=0.49\textwidth]{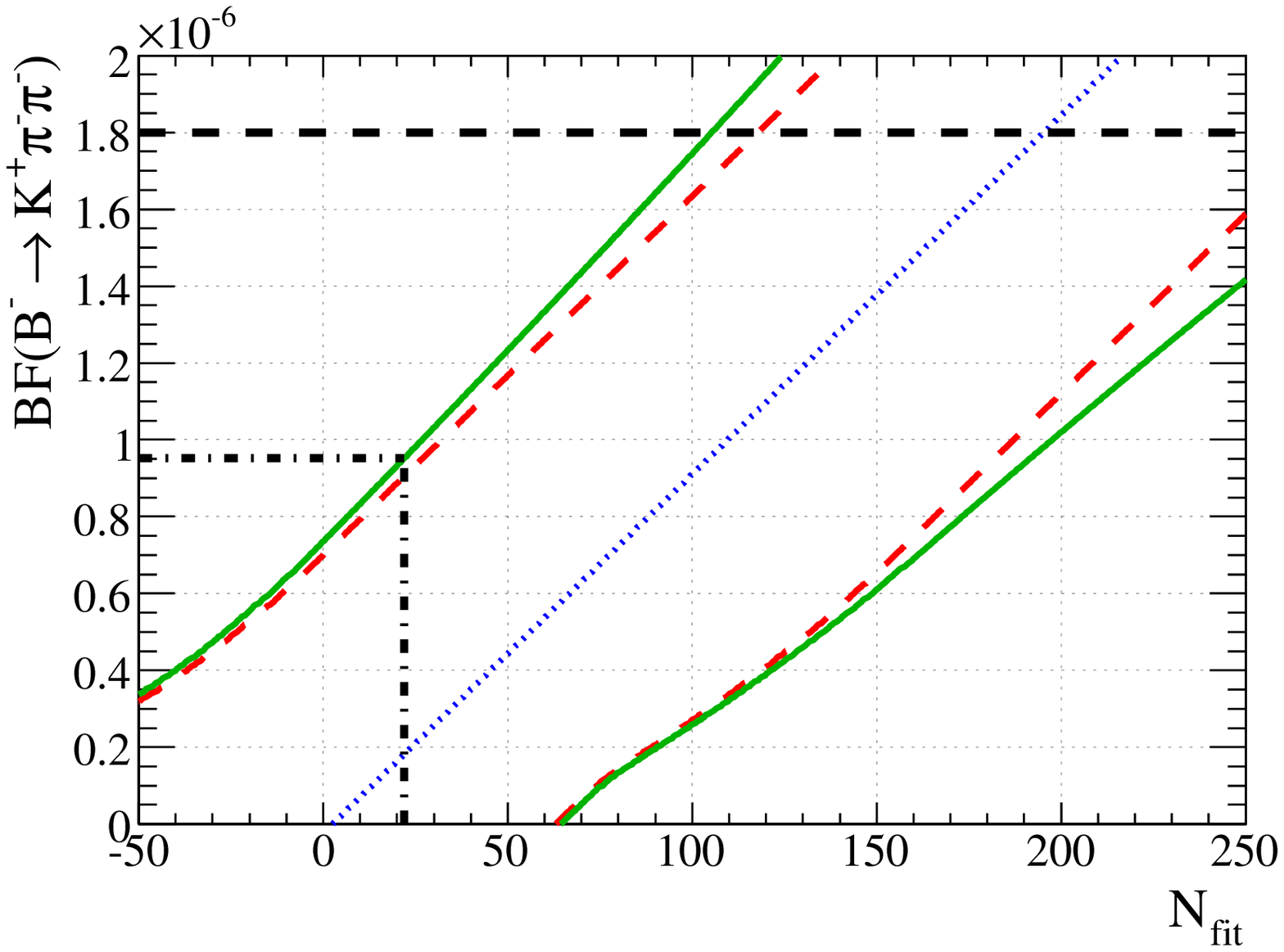}
\includegraphics[width=0.49\textwidth]{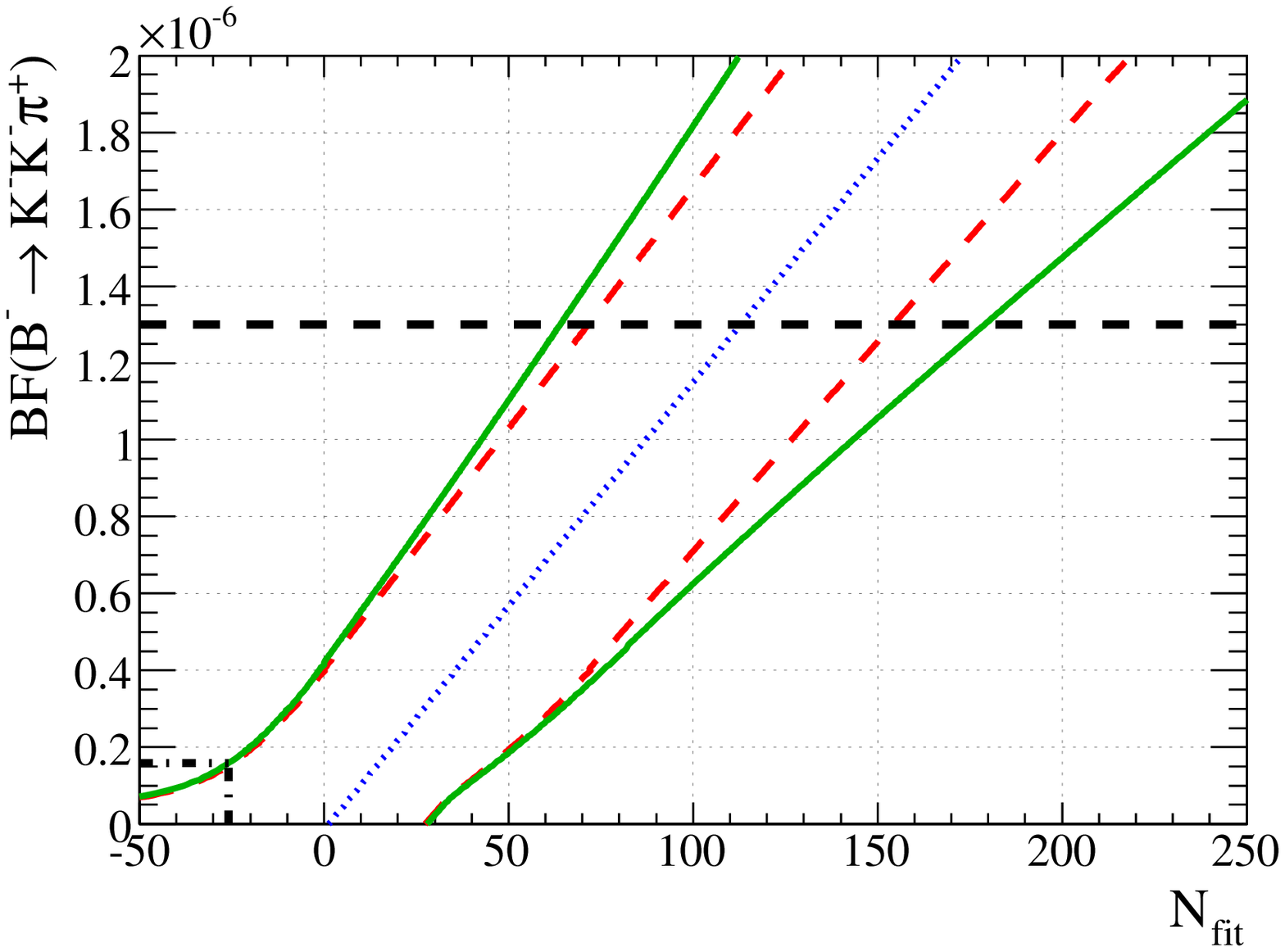}
\caption{
Construction of the confidence region bands.
The left (right) plot is for \BtoKPPSMSneg\ (\BtoKKPSMSneg).
In each figure the blue dotted line shows the expected central value of
$N_{\rm fit}$ as a function of the true branching fraction, the green solid
(red dashed) lines show the 90\% confidence level upper and lower limits
including statistical and systematic errors (statistical errors only),
the black dashed horizontal line marks the position of the previous upper
limit~\cite{Aubert:2003xz}, and the black dash-dotted lines indicate the
results of this study.
}
\label{fig:limits}
\end{figure*}

To obtain 90\% confidence level upper limits on the branching fractions, 
we use the frequentist approach of Feldman and Cousins~\cite{Feldman:1997qc}.
We determine 90\% confidence region bands that
relate the true values of the branching fractions to the measured numbers
of signal events. 
These bands are constructed using the results of MC studies that account
for relevant biases in the fit procedure and include systematic
uncertainties.  The construction of the confidence region bands is shown in
\shortfigref{limits}.
The 90\% confidence level upper limits are found to be 
${\cal B}(\BtoKPPSMSneg) < \kppUL$ and
${\cal B}(\BtoKKPSMSneg) < \kkpUL$.
To aid comparison with other experiments, we also extract the sensitivities
${\cal B}_0$ defined as the 90\% confidence level upper limits that would be
obtained in the case of zero fitted signal yield.  The sensitivities are
${\cal B}_0(\BtoKPPSMSneg) < \kppSens$ and
${\cal B}_0(\BtoKKPSMSneg) < \kkpSens$.

In conclusion, we present searches for the standard model suppressed
\B\ meson decays \BtoKPPSMSneg\ and \BtoKKPSMSneg.
We do not see any evidence of these decays and obtain improved 90\%
confidence level upper limits on the branching fractions.
These results supersede those of our previous
publication~\cite{Aubert:2003xz} and can be used to constrain models of new
physics.

We are grateful for the excellent luminosity and machine conditions
provided by our \pep2\ colleagues, 
and for the substantial dedicated effort from
the computing organizations that support \babar.
The collaborating institutions wish to thank 
SLAC for its support and kind hospitality. 
This work is supported by
DOE
and NSF (USA),
NSERC (Canada),
CEA and
CNRS-IN2P3
(France),
BMBF and DFG
(Germany),
INFN (Italy),
FOM (The Netherlands),
NFR (Norway),
MES (Russia),
MEC (Spain), and
STFC (United Kingdom). 
Individuals have received support from the
Marie Curie EIF (European Union) and
the A.~P.~Sloan Foundation.

\bibliography{references}

\begin{thebibliography}{28}
\expandafter\ifx\csname natexlab\endcsname\relax\def\natexlab#1{#1}\fi
\expandafter\ifx\csname bibnamefont\endcsname\relax
  \def\bibnamefont#1{#1}\fi
\expandafter\ifx\csname bibfnamefont\endcsname\relax
  \def\bibfnamefont#1{#1}\fi
\expandafter\ifx\csname citenamefont\endcsname\relax
  \def\citenamefont#1{#1}\fi
\expandafter\ifx\csname url\endcsname\relax
  \def\url#1{\texttt{#1}}\fi
\expandafter\ifx\csname urlprefix\endcsname\relax\def\urlprefix{URL }\fi
\providecommand{\bibinfo}[2]{#2}
\providecommand{\eprint}[2][]{\url{#2}}

\bibitem[{\citenamefont{Cabibbo}(1963)}]{Cabibbo:1963yz}
\bibinfo{author}{\bibfnamefont{N.}~\bibnamefont{Cabibbo}},
  \bibinfo{journal}{Phys. Rev. Lett.} \textbf{\bibinfo{volume}{10}},
  \bibinfo{pages}{531} (\bibinfo{year}{1963}).

\bibitem[{\citenamefont{Kobayashi and Maskawa}(1973)}]{Kobayashi:1973fv}
\bibinfo{author}{\bibfnamefont{M.}~\bibnamefont{Kobayashi}} \bibnamefont{and}
  \bibinfo{author}{\bibfnamefont{T.}~\bibnamefont{Maskawa}},
  \bibinfo{journal}{Prog. Theor. Phys.} \textbf{\bibinfo{volume}{49}},
  \bibinfo{pages}{652} (\bibinfo{year}{1973}).

\bibitem[{\citenamefont{Huitu et~al.}(1998)\citenamefont{Huitu, Zhang, Lu, and
  Singer}}]{Huitu:1998vn}
\bibinfo{author}{\bibfnamefont{K.}~\bibnamefont{Huitu}},
  \bibinfo{author}{\bibfnamefont{D.~X.} \bibnamefont{Zhang}},
  \bibinfo{author}{\bibfnamefont{C.~D.} \bibnamefont{Lu}}, \bibnamefont{and}
  \bibinfo{author}{\bibfnamefont{P.}~\bibnamefont{Singer}},
  \bibinfo{journal}{Phys. Rev. Lett.} \textbf{\bibinfo{volume}{81}},
  \bibinfo{pages}{4313} (\bibinfo{year}{1998}).

\bibitem[{\citenamefont{Fajfer et~al.}(2006)\citenamefont{Fajfer, Kamenik, and
  Kosnik}}]{Fajfer:2006av}
\bibinfo{author}{\bibfnamefont{S.}~\bibnamefont{Fajfer}},
  \bibinfo{author}{\bibfnamefont{J.~F.} \bibnamefont{Kamenik}},
  \bibnamefont{and} \bibinfo{author}{\bibfnamefont{N.}~\bibnamefont{Kosnik}},
  \bibinfo{journal}{Phys. Rev.} \textbf{\bibinfo{volume}{D74}},
  \bibinfo{pages}{034027} (\bibinfo{year}{2006}).

\bibitem[{\citenamefont{Fajfer and Singer}(2000)}]{Fajfer:2000ny}
\bibinfo{author}{\bibfnamefont{S.}~\bibnamefont{Fajfer}} \bibnamefont{and}
  \bibinfo{author}{\bibfnamefont{P.}~\bibnamefont{Singer}},
  \bibinfo{journal}{Phys. Rev.} \textbf{\bibinfo{volume}{D62}},
  \bibinfo{pages}{117702} (\bibinfo{year}{2000}).

\bibitem[{\citenamefont{Chun and Lee}(2003)}]{Chun:2003rg}
\bibinfo{author}{\bibfnamefont{E.~J.} \bibnamefont{Chun}} \bibnamefont{and}
  \bibinfo{author}{\bibfnamefont{J.~S.} \bibnamefont{Lee}}
  (\bibinfo{year}{2003}), \eprint{hep-ph/0307108}.

\bibitem[{\citenamefont{Browder et~al.}(2008)\citenamefont{Browder, Gershon,
  Pirjol, Soni, and Zupan}}]{Browder:2008em}
\bibinfo{author}{\bibfnamefont{T.~E.} \bibnamefont{Browder}},
  \bibinfo{author}{\bibfnamefont{T.}~\bibnamefont{Gershon}},
  \bibinfo{author}{\bibfnamefont{D.}~\bibnamefont{Pirjol}},
  \bibinfo{author}{\bibfnamefont{A.}~\bibnamefont{Soni}}, \bibnamefont{and}
  \bibinfo{author}{\bibfnamefont{J.}~\bibnamefont{Zupan}}
  (\bibinfo{year}{2008}), \eprint{arXiv:0802.3201 [hep-ph]}.

\bibitem[{\citenamefont{Bergfeld et~al.}(1996)}]{Bergfeld:1996dd}
\bibinfo{author}{\bibfnamefont{T.}~\bibnamefont{Bergfeld}} \bibnamefont{et~al.}
  (\bibinfo{collaboration}{CLEO Collaboration}), \bibinfo{journal}{Phys. Rev.
  Lett.} \textbf{\bibinfo{volume}{77}}, \bibinfo{pages}{4503}
  (\bibinfo{year}{1996}).

\bibitem[{\citenamefont{Abbiendi et~al.}(2000)}]{Abbiendi:1999st}
\bibinfo{author}{\bibfnamefont{G.}~\bibnamefont{Abbiendi}} \bibnamefont{et~al.}
  (\bibinfo{collaboration}{OPAL Collaboration}), \bibinfo{journal}{Phys. Lett.}
  \textbf{\bibinfo{volume}{B476}}, \bibinfo{pages}{233} (\bibinfo{year}{2000}).

\bibitem[{\citenamefont{Aubert et~al.}(2003)}]{Aubert:2003xz}
\bibinfo{author}{\bibfnamefont{B.}~\bibnamefont{Aubert}} \bibnamefont{et~al.}
  (\bibinfo{collaboration}{\babar\ Collaboration}), \bibinfo{journal}{Phys.
  Rev. Lett.} \textbf{\bibinfo{volume}{91}}, \bibinfo{pages}{051801}
  (\bibinfo{year}{2003}).

\bibitem[{\citenamefont{Garmash et~al.}(2004)}]{Garmash:2003er}
\bibinfo{author}{\bibfnamefont{A.}~\bibnamefont{Garmash}} \bibnamefont{et~al.}
  (\bibinfo{collaboration}{Belle Collaboration}), \bibinfo{journal}{Phys. Rev.}
  \textbf{\bibinfo{volume}{D69}}, \bibinfo{pages}{012001}
  (\bibinfo{year}{2004}).

\bibitem[{\citenamefont{Aubert et~al.}(2006)}]{Aubert:2006aw}
\bibinfo{author}{\bibfnamefont{B.}~\bibnamefont{Aubert}} \bibnamefont{et~al.}
  (\bibinfo{collaboration}{\babar\ Collaboration}), \bibinfo{journal}{Phys.
  Rev.} \textbf{\bibinfo{volume}{D74}}, \bibinfo{pages}{051104}
  (\bibinfo{year}{2006}).

\bibitem[{\citenamefont{Aubert et~al.}(2007{\natexlab{a}})}]{Aubert:2007fm}
\bibinfo{author}{\bibfnamefont{B.}~\bibnamefont{Aubert}} \bibnamefont{et~al.}
  (\bibinfo{collaboration}{\babar\ Collaboration}), \bibinfo{journal}{Phys.
  Rev.} \textbf{\bibinfo{volume}{D76}}, \bibinfo{pages}{071104}
  (\bibinfo{year}{2007}{\natexlab{a}}).

\bibitem[{\citenamefont{Aubert et~al.}(2008{\natexlab{a}})}]{Aubert:2007xc}
\bibinfo{author}{\bibfnamefont{B.}~\bibnamefont{Aubert}} \bibnamefont{et~al.}
  (\bibinfo{collaboration}{\babar\ Collaboration}), \bibinfo{journal}{Phys.
  Rev. Lett.} \textbf{\bibinfo{volume}{100}}, \bibinfo{pages}{081801}
  (\bibinfo{year}{2008}{\natexlab{a}}).

\bibitem[{\citenamefont{Kozanecki}(2000)}]{pep2}
\bibinfo{author}{\bibfnamefont{W.}~\bibnamefont{Kozanecki}},
  \bibinfo{journal}{Nucl. Instrum. Methods Phys. Res., Sect. A}
  \textbf{\bibinfo{volume}{446}}, \bibinfo{pages}{59} (\bibinfo{year}{2000}).

\bibitem[{\citenamefont{Aubert et~al.}(2002)}]{Aubert:2001tu}
\bibinfo{author}{\bibfnamefont{B.}~\bibnamefont{Aubert}} \bibnamefont{et~al.}
  (\bibinfo{collaboration}{\babar\ Collaboration}), \bibinfo{journal}{Nucl.
  Instrum. Methods Phys. Res., Sect. A} \textbf{\bibinfo{volume}{479}},
  \bibinfo{pages}{1} (\bibinfo{year}{2002}).

\bibitem[{\citenamefont{Aubert et~al.}(2005)}]{Aubert:2004zt}
\bibinfo{author}{\bibfnamefont{B.}~\bibnamefont{Aubert}} \bibnamefont{et~al.}
  (\bibinfo{collaboration}{\babar\ Collaboration}), \bibinfo{journal}{Phys.
  Rev. Lett.} \textbf{\bibinfo{volume}{94}}, \bibinfo{pages}{161803}
  (\bibinfo{year}{2005}).

\bibitem[{\citenamefont{Agostinelli et~al.}(2003)}]{Agostinelli:2002hh}
\bibinfo{author}{\bibfnamefont{S.}~\bibnamefont{Agostinelli}}
  \bibnamefont{et~al.} (\bibinfo{collaboration}{GEANT4 Collaboration}),
  \bibinfo{journal}{Nucl. Instrum. Methods Phys. Res., Sect. A}
  \textbf{\bibinfo{volume}{506}}, \bibinfo{pages}{250} (\bibinfo{year}{2003}).

\bibitem[{\citenamefont{Lange}(2001)}]{Lange:2001uf}
\bibinfo{author}{\bibfnamefont{D.~J.} \bibnamefont{Lange}},
  \bibinfo{journal}{Nucl. Instrum. Methods Phys. Res., Sect. A}
  \textbf{\bibinfo{volume}{462}}, \bibinfo{pages}{152} (\bibinfo{year}{2001}).

\bibitem[{\citenamefont{Oreglia}(1980)}]{Oreglia:1980cs}
\bibinfo{author}{\bibfnamefont{M.}~\bibnamefont{Oreglia}}
  (\bibinfo{year}{1980}), \bibinfo{note}{{P}hD Thesis, {S}LAC-0236, Appendix
  D}.

\bibitem[{\citenamefont{Gaiser}(1982)}]{Gaiser:1982yw}
\bibinfo{author}{\bibfnamefont{J.}~\bibnamefont{Gaiser}}
  (\bibinfo{year}{1982}), \bibinfo{note}{{P}hD Thesis, {S}LAC-0255, Appendix
  F}.

\bibitem[{\citenamefont{Skwarnicki}(1986)}]{Skwarnicki:1986xj}
\bibinfo{author}{\bibfnamefont{T.}~\bibnamefont{Skwarnicki}}
  (\bibinfo{year}{1986}), \bibinfo{note}{{P}hD Thesis, {D}ESY-F31-86-02,
  Appendix E}.

\bibitem[{\citenamefont{Albrecht et~al.}(1990)}]{Albrecht:1990am}
\bibinfo{author}{\bibfnamefont{H.}~\bibnamefont{Albrecht}} \bibnamefont{et~al.}
  (\bibinfo{collaboration}{ARGUS Collaboration}), \bibinfo{journal}{Z. Phys.}
  \textbf{\bibinfo{volume}{C48}}, \bibinfo{pages}{543} (\bibinfo{year}{1990}).

\bibitem[{\citenamefont{Aston et~al.}(1988)}]{Aston:1987ir}
\bibinfo{author}{\bibfnamefont{D.}~\bibnamefont{Aston}} \bibnamefont{et~al.},
  \bibinfo{journal}{Nucl. Phys.} \textbf{\bibinfo{volume}{B296}},
  \bibinfo{pages}{493} (\bibinfo{year}{1988}).

\bibitem[{\citenamefont{Aubert et~al.}(2008{\natexlab{b}})}]{Aubert:2008bj}
\bibinfo{author}{\bibfnamefont{B.}~\bibnamefont{Aubert}} \bibnamefont{et~al.}
  (\bibinfo{collaboration}{\babar\ Collaboration}), \bibinfo{journal}{Phys.
  Rev.} \textbf{\bibinfo{volume}{D78}}, \bibinfo{pages}{012004}
  (\bibinfo{year}{2008}{\natexlab{b}}).

\bibitem[{\citenamefont{Pivk and Le~Diberder}(2005)}]{Pivk:2004ty}
\bibinfo{author}{\bibfnamefont{M.}~\bibnamefont{Pivk}} \bibnamefont{and}
  \bibinfo{author}{\bibfnamefont{F.~R.} \bibnamefont{Le~Diberder}},
  \bibinfo{journal}{Nucl. Instrum. Methods Phys. Res., Sect. A}
  \textbf{\bibinfo{volume}{555}}, \bibinfo{pages}{356} (\bibinfo{year}{2005}).

\bibitem[{\citenamefont{Aubert et~al.}(2007{\natexlab{b}})}]{Aubert:2007xb}
\bibinfo{author}{\bibfnamefont{B.}~\bibnamefont{Aubert}} \bibnamefont{et~al.}
  (\bibinfo{collaboration}{\babar\ Collaboration}), \bibinfo{journal}{Phys.
  Rev. Lett.} \textbf{\bibinfo{volume}{99}}, \bibinfo{pages}{221801}
  (\bibinfo{year}{2007}{\natexlab{b}}).

\bibitem[{\citenamefont{Feldman and Cousins}(1998)}]{Feldman:1997qc}
\bibinfo{author}{\bibfnamefont{G.~J.} \bibnamefont{Feldman}} \bibnamefont{and}
  \bibinfo{author}{\bibfnamefont{R.~D.} \bibnamefont{Cousins}},
  \bibinfo{journal}{Phys. Rev.} \textbf{\bibinfo{volume}{D57}},
  \bibinfo{pages}{3873} (\bibinfo{year}{1998}).

\end{thebibliography}
\bibliographystyle{apsrev}

\end{document}